\documentclass[aps,showpacs,preprintnumbers,nofootinbib]{revtex4}

\usepackage{bbm}                                                  
\usepackage{bm}
\usepackage{graphicx}

\newcommand{\nc}{\newcommand}
\nc{\beq}{\begin{equation}}  \nc{\eeq}{\end{equation}}
\nc{\bea}{\begin{eqnarray}}  \nc{\eea}{\end{eqnarray}}

\def\zBB{{\mathbbm Z}}
\def\mBB{{\mathbbm M}}
\def\rBB{{\mathbbm R}}
\def\vBB{{\mathbbm V}}
\def\BB{{\bf B}}
\def\CC{{\bf C}}
\def\yy{{\bf y}}
\def\acal{{\cal A}}
\def\bcal{{\cal B}}
\def\gcal{{\cal G}}
\def\lcal{{\cal L}}
\def\ssb{spontaneous symmetry breaking}
\def\vev{vacuum expectation value}
\def\su#1{{SU(#1)}}
\def\ui{U(1)}
\def\mati{{\mathbbm1}}
\def\up#1{^{\left( #1 \right) }}
\def\alpbf{{\bm\alpha}}		
\def\betbf{{\bm\beta}}		
\def\vevof#1{\left\langle#1\right\rangle}

\begin{document}

\preprint{IFT-06-20\cr
UCRHEP-T421}

\title{Strategies and obstacles in constructing realistic higher-dimensional models}


\author{Bohdan Grzadkowski}
\address{Institute of Theoretical Physics,  Warsaw University,
Ho\.za 69, PL-00-681 Warsaw, Poland}

\author{Jos\'e Wudka}
\address{Department of Physics, University of California,
Riverside CA 92521-0413, USA}

\begin{abstract}
We discuss several aspects of higher dimensional models
that contain bulk gauge and fermion fields only.
In particular we argue that non-standard boundary conditions involving 
charge-conjugate fermion fields offer attractive model building possibilities.
We also discuss a no-go theorem 
for 5-dimensional models which severely limits their phenomenological relevance.
\end{abstract}

\maketitle


\section{Characteristics of the models considered}

We consider a space time of the form~\cite{Quiros:2003gg}
$\mBB \times \BB$ ($\mBB=$ usual 4-d Minkowski space) and \BB\ a space-like $d-4$ dimensional 
manifold. We assume that a discrete group $\gcal$ acts on \BB, $ y \to q(y)\in\BB$, and 
replace $ \BB \to \BB/\gcal$
which is compact. If we expand
the fields in some complete set of functions, the modes that do not depend on the \BB\ coordinates 
will correspond to the light excitations.
As an example take $ d=5$, $\BB = \rBB$ and $ \gcal=\{ E,~I,~T\} $
\beq 
E:~y \to y ,~~ I:~y \to - y ,~~ T:~y \to y+L; \qquad [T,I]=0,~~ I^2 = E
\eeq
$ \BB/\gcal$ denotes an orbilfod $S^1/Z_2$.
Fourier expanding the 
fields~\footnote{Minkowski-space coordinates are suppressed to 
minimize cluttering.}
\bea
\phi(y) &=& \sum_{n\in \zBB} \phi_{n+} \; \cos(2\pi n y/L) + 
\sum_{n\in \zBB} \phi_{n-} \; \sin(2\pi n y/L)
\eea
The kinetic energy contains a term $ \propto \partial_y^2 $ that generates a mass
$ \sim |2\pi n/L| $, so that all the $n\not=0$ modes are heavy. Light modes, $\phi_{0+}$ 
are associated with $y-$independent functions.

In these theories it is possible to recast the
hierarchy problem in terms of the compactification scale\cite{Hosotani:1988bm}:
consider~\footnote{No brane kinetic 
terms~\cite{Burdman:2002se}, 5-d bulk scalars, 
nor anomalous $U(1)$~\cite{Antoniadis:2001cv} will be considered hereafter.}
$ \lcal = - F^2/4 + 
\bar\Psi ( i \not\!\partial -M ) \Psi $ in 5d, then 
\beq
H \sim \int_0^L A_{N=4}(x,y) dy,
\label{eq:h}
\eeq
generates 4-d scalar excitations ($H \not=0$ 
if we impose appropriate behavior of $ A_4$ under $ \gcal$).
Radiative corrections generate a non-trivial effective potential $ V_{\rm eff}(H)$, 
which is {\em finite} (up to field independent constant) 
and calculable. The reason is that all divergences 
in the full 5-d theory are
associated with local operators while none of the terms in $ V_{\rm eff}(H)$
can be local since $H$ itself is not. The $H$ mass
is usually $O(1/L)$ and is not driven to an UV cutoff by radiative corrections.

\section{Non-standard boundary conditions}

\begin{figure}[h]
\includegraphics[bb=72 72 540 720, height=8in]{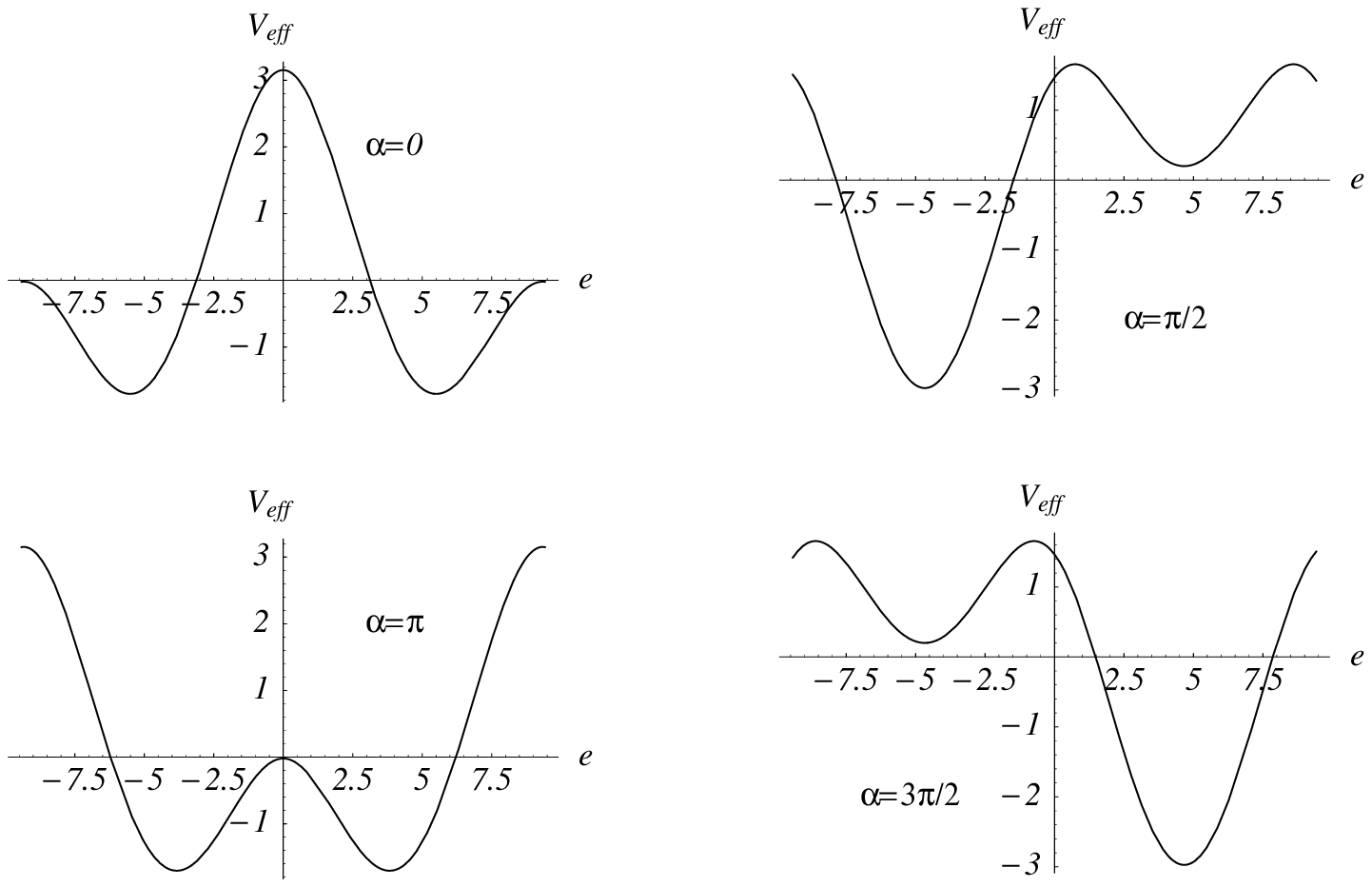}
  \caption{Effective potential for an Abelian model with non-standard
boundary conditions when two fermions of charges $2/3$ and $-1/3$ are
present, for the remaining parameters see \cite{Grzadkowski:2006tp}.}
\label{fig:fig}
\end{figure}

We will be interested in models without fundamental scalars, 
where $H$ (eq. \ref{eq:h}) drives \ssb\ (SSB) in the light theory. 
For example, in a 5-d Abelian model with
$ A_4(y+L) = A_4(-y) = A_4(y) $,
invariance under $\gcal$ requires $ A_\mu(-y) = - A_\mu(y) $,
while for the fermions one usually {\em assumes}
$ \psi(-y) = e^{i \beta} \gamma_5 \psi(y) $. But in this case
$
\bar\psi (\not\!\partial + i g \not\! A )\psi
\to 
\bar\psi (\not\!\partial+ i(- g) \not\!A )\psi 
$, which
is {\em incompatible with $g\not=0$}. Since the $A_N$ behavior under $\gcal$ is necessary
for SSB, we must modify that for the $\psi$, and the sign change in $g$
suggests we include the charge conjugate field $ \psi^c$. 
This illustrated by the following choices for an Abelian model \cite{Grzadkowski:2005rz}:
\beq
\begin{array}{|l|l|} \hline
\hfill Standard	\hfill & \hfill Non-standard \hfill \cr \hline
\psi(y) = \gamma_5 \psi(-y) = e^{-i \alpha} \psi(y+L)& 
\psi(y) = \gamma_5 \psi^c(-y)  = e^{-i \alpha} \psi(y+L)\cr
A^\mu(y) = A^\mu(-y) = A^\mu(y+L) & A^\mu(y) = - A^\mu(-y) = A^\mu(y+L) \cr
A^4(y)   = - A^4(-y) =   A^4(y+L) & A^4(y) =     + A^4(-y) =   A^4(y+L) \cr \hline
\end{array}
\eeq
The light spectrum contains a vector (scalar) for the (non) standard boundary conditions
and a chiral fermion when $ \alpha \simeq0 $ (both cases). No symmetry
protects the light-scalar mass and radiative corrections will, in general,
generate a non-trivial effective potential that may lead to SSB, see Fig.\ref{fig:fig} for
$\alpha=0,\pi$. This
does happen, but it requires 2 fermion species
with different charges (see fig. \ref{fig:fig}); the scalar then receives a mass and \vev\
(VEV) $\sim 1/L$:
after one loop effects are taken into account the model has no massless excitations.
Expanding in a Fourier series the fields take the form 
\beq
 A^\mu(y) = \sum_n A^\mu_n\; \sin(\omega_n y) ; \quad 
 A^4(y) = \sum_n A_n^4\; \cos( \omega_n y ); \quad 
 \psi(y) = \sum_n e^{i \tilde\omega_n } \left( \begin{array}{c} 
 \varepsilon \phi_n^* \cr \phi_n \end{array} \right) ,
\eeq
where $\omega_n = 2\pi n/L,~
\tilde\omega_n = \omega_n + \alpha/L$ and $\varepsilon \equiv i \sigma_2$ is the $2\times 2$ antisymmetric matrix.
In particular the $\phi_n$ receive a Majorana mass terms of the form
$ \sum_n \tilde\omega_n \; \phi_n^T \varepsilon \phi_n$, and may prove useful
when constructing the neutrino sector of the electroweak theory.

For $ \gcal = \{ E, T\}$ (no ``orbifolding'')
the model will contain light vector-like fermions, a vector boson and a light scalar
with SSB when $ \ge2$ fermion flavors of different charges are present in which case
both fermions and scalars receive $O(1/L)$ masses. It is worth to mention that in
this model (for $\alpha=0,\pi$) spontaneous CP violation takes place \cite{Grzadkowski:2006tp}.

In more complicated groups one can mix boundary conditions
insuring that the $A^a_N$ contain 4-dimensional light vectors
for some gauge indices $a$, and light scalars for others. 
For example in a 5-d $\su3$ theory we can choose
$A^{a=1,2,3,8}_{N\not=4}$ and $ A^{a=4,5,6,7}_{N=4} $ to be even and periodic in $y$.
The light modes of the first set will be the gauge boson of a 4d $\su2\times\ui$ theory
while those of the second set form an $\su2$ doublet (4d scalar) that can receive a VEV
through 1-loop effective potential. The $\rho$ parameter satisfies
$\rho_{\rm tree} = 1 $ and a careful choice of the fermionic content 
leads to VEV$\ll 1/L$; unfortunately the weak mixing
angle differs widely form the observed value: 
$\sin^2 \theta_W = 3/4$.

The general case\cite{Grzadkowski:2005rz} is best described in terms of a double spinor
$\chi^T = (\psi^T, -\psi^{c\;T})$:
\beq
\chi(y) = \acal \chi(y-L) = -\gamma_5 \bcal \chi(-y) 
\quad A_a^N(y) = \vBB_{ab} A_b^N(y-L) = (-1)^{\delta_{N,4}} \tilde\vBB_{ab} A_b^N(-y)
\eeq
where $\acal,\bcal$ are unitary, $ \vBB,\tilde\vBB$ orthogonal and satisfy
\beq
\mati = -\varepsilon \acal \varepsilon \acal^T = \varepsilon \bcal \varepsilon \bcal^T; \quad
f_{a'b'c'} = f_{abc} \vBB_{a a'} \vBB_{b b'} \vBB_{c c'} =
f_{abc} \tilde\vBB_{a a'}  \tilde\vBB_{b b'}  \tilde\vBB_{c c'}
\eeq
The first relation insures that $ \chi$ and $ \psi $ have the same
number of degrees of freedom; the second implies
that $\vBB,~\tilde\vBB$ are automorphisms of the gauge group.

The  light modes (denoted by a $\up0$ superscript) then satisfy
\beq
\chi\up0 = \acal \chi\up0 = -\gamma_5 \bcal \chi\up0 \qquad
A_N\up0 = \vBB A_N\up0 = (-1)^{\delta_{N4}} \tilde\vBB A_N\up0
\eeq
By appropriate choice of group and fermion content one can have
Dirac and/or Majorana fermion masses, as well as
radiatively-induced SSB for the $A_4\up0$.

\section{Constraints on realistic models}

The best case scenario would
correspond to one which experiences two stages of symmetry breaking:
$ G \to \su3\times\su2\times\ui \to\ui$
The first generated by the choice of $ \vBB,~\acal,$ etc. and the
second by SSB through the scalars $H$. For these
models to be phenomenologically viable this should lead to 
$ \rho_{\rm tree} = 1$ and $\sin^2\theta_W \simeq 1/4$.
Since all couplings are 
specified by the group structure, it is possible to determine 
in full generality under which conditions these constraints are satisfied.
Demanding that there be exclusively SM vector bosons 
implies the model can contain only one light scalar multiplet
of isospin $I_{\rm max}$, 
associated with a root $ \betbf$, whose component of $z$-isospin $I$ 
gets a non-zero VEV.
The gauge vectors can be expanded as 
\beq
A_\mu = E_{\alpbf} W_\mu^+ +E_{-\alpbf} W_\mu^- + \hat\alpbf \cdot \CC ~W_\mu^0 + \hat\yy\cdot\CC ~ B_\mu + \cdots; \qquad
A_4 = \phi E_\betbf + \phi^\dagger E_{-\betbf} 
\eeq
(Cartan and root generators are denoted by $C_i $ and $E_\alpbf$
respectively) where $ \alpbf$ is a root not parallel to $\betbf$ and
$ \hat\yy = (\betbf - (\hat\alpbf\cdot\betbf)\hat\alpbf)/|(\betbf - (\hat\alpbf\cdot\betbf)\hat\alpbf) | $.
The masses are generated by 
\beq
\lcal = (\vevof H^2/2)
\left\{ |\alpbf|^2 \left[I_{\rm max}\left(I_{\rm max}+1\right)
-I^2 \right]W^+\cdot W^-
+ \left( \hat\alpbf\cdot\betbf W^0+\hat\yy\cdot\betbf B\right)^2
\right\}
\eeq
so
$
\sin^2\theta_W = 1 - (\hat\alpbf\cdot\hat\betbf)^2; ~
\rho_{\rm tree} = I_{\rm max} \left(I_{\rm max}+1\right)/(2I^2)- 1/2
$. Standard Lie algebra properties imply
$ 2|I| = 4 (\hat\alpbf \cdot\hat\betbf)^2 = 0,1,2,3,4 $, which
combined with
$\rho_{\rm tree} =1 $,
imply
$
\sin^2 \theta_W = 3/4,~0
$: 
{\em any 5-d model 
without scalars, kinetic brane terms or anomalous $U(1)$ is
unacceptable} \cite{Grzadkowski:2006tp}.

\section{Conclusions}
We have discussed non-standard boundary conditions involving 
charge-conjugate fermion fields. They offer attractive model 
building possibilities. In particular the Abelian example shows that
the non-standard fermionic boundary conditions are mandatory when
the $U(1)$ gauge symmetry is supposed to be broken in the light sector
of 4d effective theory. We have also presented a no-go theorem 
for 5d models which severely limits their phenomenological relevance.


\begin{acknowledgments}
 Work supported in part by the Ministry of Science and Higher Education (Poland) 
in years 2004-6 and 2006-8 as research projects
1~P03B~078~26 and N202~176~31/3844, respectively, and by the European Community 
under project MTKD-CT-2005-029466; and by funds provided 
by the U.S. Department of Energy under grant No.~DE-FG03-94ER40837.

\end{acknowledgments}




\begin{thebibliography}{9}

\bibitem{Quiros:2003gg}
 For a review see: M.~Quiros,
 ``New ideas in symmetry breaking,''
  arXiv:hep-ph/0302189.

\bibitem{Hosotani:1988bm}
  Y.~Hosotani,
  Annals Phys.\  {\bf 190}, 233 (1989).

\bibitem{Grzadkowski:2005rz}
  B.~Grzadkowski and J.~Wudka,
  Phys.\ Rev.\ D {\bf 72}, 125012 (2005)
  [arXiv:hep-ph/0501238];
  Acta Phys.\ Polon.\ B {\bf 36}, 3523 (2005)
  [arXiv:hep-ph/0511139].
  
\bibitem{Burdman:2002se}
  G.~Burdman and Y.~Nomura,
  Nucl.\ Phys.\ B {\bf 656}, 3 (2003)
  [arXiv:hep-ph/0210257];
K.~Agashe, R.~Contino and A.~Pomarol,
Nucl.\ Phys.\ B {\bf 719}, 165 (2005)
[arXiv:hep-ph/0412089].

\bibitem{Antoniadis:2001cv}
  I.~Antoniadis, K.~Benakli and M.~Quiros,
  New J.\ Phys.\  {\bf 3}, 20 (2001)
  [arXiv:hep-th/0108005];
  C.~A.~Scrucca, M.~Serone and L.~Silvestrini,
  Nucl.\ Phys.\ B {\bf 669}, 128 (2003)
  [arXiv:hep-ph/0304220];
  G.~Cacciapaglia, C.~Csaki and S.~C.~Park,
  arXiv:hep-ph/0510366.

\bibitem{Grzadkowski:2006tp}
  B.~Grzadkowski and J.~Wudka,
  arXiv:hep-ph/0604225.

\end{thebibliography}
\end{document}